# Open access institutional and news media tweet dataset for COVID-19 social science research

**Jingyuan Yu**

Department of Social Psychology, Universitat Autònoma de Barcelona

Jingyuan.yu@e-campus.uab.cat

## Abstract

As COVID-19 quickly became one of the most concerned global crisis, the demand for data in academic research is also increasing. Currently, there are several open access Twitter datasets, but none of them is dedicated to the institutional and news media Twitter data collection. To fill this blank, we retrieved data from 69 institutional/news media Twitter accounts, 17 of them were related to government and international organizations, 52 of them were news media across North America, Europe and Asia. We believe our open access data can provide researchers more availability to conduct social science research.

## Introduction

COVID-19 was announced as pandemic by WHO on 11 Mar [1], being the only pandemic in the past 10 years (the last one was 2009 swing flu), it was first detected as an unknown pneumonia in Wuhan (Hubei, China). On April $1_{st}$, 2020, John Hopkins University Coronavirus Resource Center [2], reported that the ongoing epidemic has infected 938,373 people, the death toll has already reached 47,272, given the highly contagious nature of the virus as well as the significant mortality rate, numerous governments have announced national lockdown. The impact of COVID-19 on world economy and politics is unprecedented in modern time.

On the past Ebola epidemic crisis, scholars found the importance of using Twitter data to do social science research [3], [4], many of them use this microblog data as social indicators to analyze the effect of epidemic outbreak on public concerns [5], health information needs and health seeking behavior [6], and public response to policy makers [7] etc. Current open access COVID-19 Twitter data were mainly collected by keywords, such as coronavirus, Covid-19 etc [8], [9], none of the them is dedicated to government/news media tweet collection. Given that our retrieval targets are policy makers and news source, we believe our dataset can provide scholars more valuable data to conduct social science research in related fields, such as crisis communication, public relation etc.

## Data collection strategy

We used Twitter REST API to retrieve twitter data from March 12, 2020, there are 8 collection categories: "gov tweet" (governments, international organizations etc.), "US news tweet", "UK news tweet", "Spain news tweet", "Germany news tweet", "France news tweet", "China news tweet" and "Additional news tweet" (news source added later), each of them contains various collection target (twitter account name, details see next section).

At the first time we collected the most recent 3200 tweets of every collection target before March 12, 2020, we didn't set a time limit, which implies that the date of the first tweet from each of the sources may vary and even be relatively old. We update our dataset every week (last update on April 2, 2020), after removing duplicated data by matching tweet id, we get the clean version data.

On the other hand, due to the data collection strategy we made, there may be tweets both related and unrelated to COVID-19, which opens up new possibilities for academic analysis.

## Data details

Gov tweet category contains the following accounts (Table1):

| Twitter accounts | Twitter account description | Collected Timespan (Y/M/D) | Total number of tweets |
|---|---|---|---|
| @WHO | World health organization | 2019/12/04 – 2020/04/01 | 4031 |
| @WHO_Europe | The WHO Regional Office for Europe | 2018/10/17 – 2020/04/01 | 3358 |
| @ECDC_EU | The European Centre for Disease Prevention and Control | 2016/06/16 – 2020/04/01 | 3466 |
| @EU_Commission | European Commission | 2019/04/04 – 2020/04/01 | 3477 |
| @CDCgov | CDC's official Twitter source for daily credible health & safety updates from Centers for Disease Control & Prevention (US) | 2018/12/18 – 2020/04/01 | 3392 |
| @CDCemergency | The handle for CDC's Center for Preparedness and Response (CPR) (US) | 2017/10/01 – 2020/04/01 | 3336 |
| @CDCGlobal | CDC global health (US) | 2017/09/07 – 2020/04/01 | 3300 |
| @HHSGov | Department of Health & Human Services (HHS) (US) | 2019/02/04 – 2020/04/01 | 3381 |
| @DHSCgovuk | Department of Health and Social Care (UK) | 2019/02/08 – 2020/04/01 | 3743 |
| @PHE_uk | Official feed of Public Health England (PHE) | 2019/02/04 – 2020/04/01 | 3401 |
| @MinSoliSante | Official account of the Ministry of Solidarity and Health (France) | 2018/12/05 – 2020/04/01 | 3588 |
| @MinisteroSalute | Official profile of the Ministry of Health (Italy) | 2015/05/22 – 2020/04/01 | 3288 |
| @BMG_Bund | Federal Ministry of Health (Germany) | 2017/10/20 – 2020/04/01 | 3735 |
| @GermanyDiplo | German Foreign Office | 2017/06/19 – 2020/04/01 | 3322 |
| @rki_de | Robert Koch Institute (Germany) | 2013/07/16 – 2020/04/01 | 2111 |

| Twitter accounts | Twitter account description | Collected Timespan (Y/M/D) | Total number of tweets |
|---|---|---|---|
| @sanidadgob | Ministry of Health (Spain) | 2019/02/08 – 2020/04/01 | 3561 |
| @SaludPublicaEs | Official account of the Government of Spain with information of interest to citizens on Public Health issues. | 2014/10/14 – 2020/04/01 | 531 |

Table 1: Data collection targets of "gov tweet" category

US news tweet category contains the following accounts (Table 2):

| Twitter accounts | Twitter account description | Collected Timespan (Y/M/D) | Total number of tweets |
|---|---|---|---|
| @nytimes | The NewYork Times | 2020/02/11 – 2020/04/01 | 5554 |
| @CNN | CNN | 2020/02/20 – 2020/04/01 | 6473 |
| @washingtonpost | The Washington Post | 2020/02/14 – 2020/04/01 | 6194 |
| @WSJ | Wall Street Journal | 2020/02/05 – 2020/04/01 | 5114 |
| @HuffPost | Huffington Post | 2020/01/23 – 2020/04/01 | 4632 |
| @latimes | Los Angeles Times | 2020/02/24 – 2020/04/01 | 6380 |
| @USATODAY | USA Today | 2020/02/07 – 2020/04/01 | 5302 |
| @BreitbartNews | Breitbart News | 2020/01/25 – 2020/04/01 | 3831 |
| @MSNBC | MSNBC | 2020/02/24 – 2020/04/01 | 4501 |

Table 2. Data collection targets of "US news tweet" category

UK news tweet category contains the following accounts (Table 3):

| Twitter accounts | Twitter account description | Collected Timespan (Y/M/D) | Total number of tweets |
|---|---|---|---|
| @Independent | The Independent | 2020/03/04 – 2020/04/01 | 11668 |
| @DailyMirror | Daily Mirror | 2020/03/04 – 2020/04/01 | 11249 |
| @Telegraph | The Telegraph | 2020/02/08 – 2020/04/01 | 5315 |
| @thetimes | The Times | 2020/01/02 – 2020/04/01 | 4214 |
| @guardian | The Guardian | 2020/02/25 – 2020/04/01 | 7592 |
| @BBCNews | BBC News (UK) | 2020/02/04 – 2020/04/01 | 5450 |
| @BBCWorld | BBC News (World) | 2019/12/29 – 2020/04/01 | 4039 |
| @BBCBreaking | BBC Breaking News | 2017/10/02 – 2020/04/01 | 3287 |

Table 3. Data collection targets of "UK news media" category

Spain news tweet category contains the following accounts (Table 4):

| Twitter accounts | Twitter account description | Collected Timespan (Y/M/D) | Total number of tweets |
|---|---|---|---|
| @LaVanguardia | La Vanguardia | 2020/02/25 – 2020/04/01 | 7775 |
| @rtve | RTVE(Spanish Radio and | 2020/02/17 – 2020/04/01 | 6293 |

| | Television Corporation) | | |
|---|---|---|---|
| @abc_es | ABC | 2020/02/24 – 2020/04/01 | 7583 |
| @elmundoes | El Mundo | 2020/02/19 – 2020/04/01 | 7049 |
| @el_pais | El Pais | 2020/02/25 – 2020/04/01 | 8253 |
| @elpaisinenglish | El Pais in English | 2019/03/27 – 2020/04/01 | 3402 |
| @elperiodico | El Periodico | 2020/02/24 – 2020/04/01 | 7273 |

Table 4. Data collection targets of "Spain news tweet" category

Germany news tweet category contains the following accounts (Table 5):

| Twitter accounts | Twitter account description | Collected Timespan (Y/M/D) | Total number of tweets |
|---|---|---|---|
| @BILD | Bild | 2020/02/19 – 2020/04/01 | 6320 |
| @DIEZEIT | Die Zeit | 2019/10/19 – 2020/04/01 | 3573 |
| @zeitonline | Zeit Online | 2020/01/25 – 2020/04/01 | 4618 |
| @derspiegel | Der Speigel | 2020/01/06 – 2020/04/01 | 4350 |
| @spiegelonline | Spiegel Online | 2019/09/13 – 2020/04/01 | 3241 |
| @faznet | Frankfurter Allgemeine | 2020/02/07 – 2020/04/01 | 5249 |
| @fr | Frankfurter Rundschau | 2019/06/05 – 2020/04/01 | 3454 |
| @SZ | Süddeutsche Zeitung | 2020/01/20 – 2020/04/01 | 4381 |
| @Tagesspiegel | Der Tagesspiegel | 2020/02/09 – 2020/04/01 | 5418 |
| @welt | Die Welt | 2020/02/23 – 2020/04/01 | 7153 |
| @tazgezwitscher | Taz news | 2019/12/27 – 2020/04/01 | 4190 |
| @FOCUS_TopNews | FOCUS Online TopNews | 2020/02/27 – 2020/04/01 | 6303 |

Table 5. Data collection targets of "Germany news tweet" category

France news tweet category contains the following accounts (Table 6):

| Twitter accounts | Twitter account description | Collected Timespan (Y/M/D) | Total number of tweets |
|---|---|---|---|
| @le_Parisien | Le parisien | 2020/02/26 – 2020/04/01 | 10193 |
| @Le_Figaro | Le Figaro | 2020/02/27 – 2020/04/01 | 9560 |
| @lemondefr | Le monde | 2020/01/31 – 2020/04/01 | 5610 |
| @LeHuffPost | Huffington Post (France) | 2020/01/17 – 2020/04/01 | 4797 |
| @France24_fr | France 24 French | 2020/01/01 – 2020/04/01 | 4189 |
| @France24_en | France 24 English | 2020/01/08 – 2020/04/01 | 4468 |
| @RFI | Radio France Internationale | 2020/02/07 – 2020/04/01 | 4764 |

Table 6. Data collection targets of "France news tweet" category

China news tweet category contains the following accounts (Table 7):

| Twitter accounts | Twitter account description | Collected Timespan | Total number of tweets |
|---|---|---|---|

| | | (Y/M/D) | |
|---|---|---|---|
| @CCTV | China Central Television | 2019/09/10 – 2020/04/01 | 3476 |
| @CGTNOfficial | China Global Television Network | 2020/02/05 – 2020/04/01 | 5262 |
| @PDChina | People's Daily | 2019/12/13 – 2020/04/01 | 4126 |
| @XHNews | Xinhua News Agency | 2020/02/06 – 2020/04/01 | 5287 |

Table 7. Data collection targets of "China news tweet" category

Additional news tweet category are later added, this category contains the following accounts (Table 8):

| Twitter accounts | Twitter account description | Collected Timespan (Y/M/D) | Total number of tweets |
|---|---|---|---|
| @repubblica | La Repubblica (Italy) | 2020/03/05 – 2020/04/01 | 12568 |
| @AJEnglish | Al Jazeera English | 2020/01/20 – 2020/04/01 | 4506 |
| @globaltimesnews | Global Times (China) | 2020/02/20 – 2020/04/20 | 6369 |
| @DeutscheWelle | DW Deutsche Welle (German wave) | 2017/12/12 – 2020/04/01 | 3288 |
| @RT_com | RT (Russia) | 2020/02/09 – 2020/04/01 | 5365 |

Table 8. Data collection targets of "Additional news tweet category"

# Data Availability

The dataset is available on Github at the following address: https://github.com/narcisoyu/Institional-and-news-media-tweet-dataset-for-COVID-19-social-science-research. Our data was collected in compliance with Twitter's official developer agreement and policy [10]. The dataset will be updated weekly, interested researchers will need to agree upon the terms of usage dictated by the chosen license.

Following Twitter official policies, we released and stored only tweet ids, as far as we know, two tools can be used to hydrate full information: Hydrator (https://github.com/DocNow/hydrator) and Twarc (https://github.com/DocNow/twarc). Interested researchers shall follow the usage instructions of the fore-mentioned tools.

# Reference

[1] WHO, "WHO Director-General's opening remarks at the media briefing on COVID-19 - 11 March 2020," 11-Mar-2020. [Online]. Available: https://www.who.int/dg/speeches/detail/who-director-general-s-opening-remarks-at-the-media-briefing-on-covid-19---11-march-2020. [Accessed: 30-Mar-2020].

[2] E. Dong, H. Du, and L. Gardner, "An interactive web-based dashboard to track COVID-19 in real time.," *Lancet. Infect. Dis.*, vol. 0, no. 0, 2020.

[3] I. C. H. Fung, Z. T. H. Tse, C. N. Cheung, A. S. Miu, and K. W. Fu, "Ebola and the social media," *The Lancet*, vol. 384, no. 9961. Lancet Publishing Group, p. 2207, 27-Dec-2014.

[4] The Lancet, "The medium and the message of Ebola," *The Lancet*, vol. 384, no. 9955. Lancet Publishing Group, p. 1641, 08-Nov-2014.

[5] A. J. Lazard, E. Scheinfeld, J. M. Bernhardt, G. B. Wilcox, and M. Suran, "Detecting themes of public concern: A text mining analysis of the Centers for Disease Control and Prevention's Ebola live Twitter chat," *Am. J. Infect. Control*, vol. 43, no. 10, pp. 1109–1111, Oct. 2015.

[6] M. Odlum and S. Yoon, "Health Information Needs and Health Seeking Behavior During the 2014-2016 Ebola Outbreak: A Twitter Content Analysis," *PLoS Curr.*, vol. 10, 2018.

[7] B. Crook, E. M. Glowacki, M. Suran, J. K. Harris, and J. M. Bernhardt, "Content Analysis of a Live CDC Twitter Chat During the 2014 Ebola Outbreak," *Commun. Res. Reports*, vol. 33, no. 4, pp. 349–355, Oct. 2016.

[8] E. Chen, K. Lerman, and E. Ferrara, "COVID-19: The First Public Coronavirus Twitter Dataset," Mar. 2020.

[9]     J. M. Banda, R. Tekumalla, and C. Gerardo, "A Twitter Dataset of 70+ million tweets related to COVID-19." Zenodo, 2020.

[10]    Twitter, "Developer Agreement and Policy – Twitter Developers," *2020*. [Online]. Available: https://developer.twitter.com/en/developer-terms/agreement-and-policy. [Accessed: 03-Apr-2020].